\documentclass[a4paper,10pt]{article}
\usepackage{dsfont, amsmath, verbatim, graphicx}
\usepackage{amssymb}
\usepackage{epsfig}
\usepackage{bbm}
\usepackage{mathtools}
\DeclarePairedDelimiter\bra{\langle}{\rvert}
\DeclarePairedDelimiter\ket{\lvert}{\rangle}
\DeclarePairedDelimiterX\braket[2]{\langle}{\rangle}{#1 \delimsize\vert #2}
\usepackage[latin1]{inputenc} % Umlaute!

\newcommand{\be}{\begin{equation}}
\newcommand{\ee}{\end{equation}}
\newcommand{\bea}{\begin{eqnarray}}
\newcommand{\eea}{\end{eqnarray}}

\newcommand{\qed}{\ensuremath{\hfill \Box}}

\newcommand{\kommentar}[1]{}

\newcommand{\forget}[1]{}

\begin{document}
\title{Completing the proof of ``Generic quantum nonlocality''}
\author{Mariami Gachechiladze and Otfried G\"uhne
\\
\small Naturwissenschaftlich-Technische Fakult\"at, Universit\"at Siegen, \\
\small Walter-Flex-Str. 3, 57068 Siegen, Germany
}
\maketitle

\begin{abstract}
In a paper by Popescu and Rohrlich 
[Phys. Lett. A {\bf 166}, 293 (1992)] a proof has been
presented showing that any pure entangled multiparticle quantum state 
violates some Bell inequality. We point out a gap in this proof, but 
we also give a construction to close this gap. It turns out that 
with some extra effort all the results from the aforementioned 
publication can be proven. Our construction shows how two-particle 
entanglement can be generated via performing local projections on
a multiparticle state.
\end{abstract}

%\pacs{}
\maketitle

\section{Introduction}
The question which quantum states violate a Bell inequality and which not is of
central importance for quantum information processing. In Ref.~\cite{pr}
it has been shown that any pure multiparticle quantum state violates a 
Bell inequality. The strategy for proving this statement was the following: 
First, one can show that for any entangled pure state on $N$ particles 
one can find projective measurements on $N-2$ particles, such that for
appropriate results of the measurements the remaining two particles
are in an entangled pure state. Then, one can apply the known fact that
any pure bipartite entangled state violates some Bell inequality~\cite{gisin}.

In this note we point out a gap in the proof presented in Ref.~\cite{pr}.
The gap concerns the part where the projective measurements on $N-2$ 
particles are made. It turns out that a certain logical step does not 
follow from the previous statements and we give an explicit 
counterexample for a conclusion drawn  at the 
critical point. Luckily it turns out, however, 
that with a significantly refined and extended argumentation the 
main statement can still be proven. Independently of the connection to
Ref.~\cite{pr} our results provide a constructive way how a two-particle 
entangled state can be generated from an $N$-particle state by performing 
local projections onto $N-2$ particles. This may be of interest for the theory
of multiparticle entanglement.

This note is organized as follows. In Section~2 we discuss the 
proof from Ref.~\cite{pr} and the problem with a Lemma used there. 
In Section~3 we present a detailed proof of the required statement
for qubits. Finally, in Section~4 we discuss the higher-dimensional
case as well as some other observations needed for the proof.

%%%%%%%%%%%%%%%%%%%%%%%%%%%%%%%%%%%%%%%%%%%%%%%%%%%%%%%
\section{Discussion of the original argument}
%%%%%%%%%%%%%%%%%%%%%%%%%%%%%%%%%%%%%%%%%%%%%%%%%%%%%%%

The gap concerns the proof of the Lemma on page 296 of Ref.~\cite{pr}. 
This lemma states that:

{\it 
Let $\ket{\psi}$ be an $N$ system entangled state. For any two of the $N$ 
systems, there exists a projection onto a direct product of state of the 
other $N-2$ systems, that leaves the two systems in an entangled state.
}
In the following we show that while the Lemma is correct, there is a gap in its original proof. Doing so, in this section  we will reformulate the proof in modern language in order to see where the problem is. For simplicity, we first consider 
only qubits.

The  proof from Ref.~\cite{pr} is a proof by contradiction, so it starts with assuming the opposite. So, orthogonal 
basis vectors $\ket{b_i} \in \{\ket{0}, \ket{1}\}$ are
considered for each qubit $i$, where the conclusion 
does not hold. That is, 
\be
\bra{b_3} \bra{b_4} \dots \bra{b_N} {\psi}\rangle = \ket{\alpha}\ket{\beta},
\ee
where the projections are carried out on the qubits $3, \dots ,N$ and the 
qubits one and two remain in the product state $\ket{\alpha}\ket{\beta}$ 
for any possible choice of the $\bra{b_3} \bra{b_4} \dots \bra{b_N}$. The 
$\bra{b_i}$ can take the values $0$ or $1$. So, the product vector will 
in general depend on this choice and it is appropriate to write this 
dependency as 
\be
 \ket{\alpha} = \ket{\alpha(b_3, \dots ,b_N)} 
\quad \mbox{ and } \quad 
 \ket{\beta}  = \ket{\beta(b_3, \dots , b_N)}.
\ee
What happens if the value of $b_3$ changes? The proof in Ref.~\cite{pr} argues convincingly 
that then not {\it both} of the $\ket{\alpha}$ and $\ket{\beta}$ can change: 
If this were the case, a projection onto the superposition 
$\bra{c_3}= \bra{b_3=0} + \bra{b_3=1}$, while keeping $\bra{b_4} \dots \bra{b_N}$
constant projects the system on the first two qubits in an entangled state. 
So, we have either
\be
\ket{\alpha} = \ket{\alpha(\circ, b_4, \dots ,b_N)} 
\quad  \mbox{ or } \quad 
 \ket{\beta}  = \ket{\beta(\circ, b_4, \dots , b_N)},
\ee
where the ``$\circ$'' indicates that $\ket{\alpha}$
or $\ket{\beta}$  {\it for the given values of $b_4, \dots ,b_N$}
does not depend on $b_3$.

The original proof continues the argument as follows: {\it Repeating the 
argument for other subspaces, we conclude that ... each index [$b_i$] 
actually appears in either $\ket{\alpha}$  or in $\ket{\beta}$ but not 
in both.} This conclusion is unwarranted. The point is that for a given 
set of $b_4,\dots, b_N$ one of the vectors (say, $\ket{\alpha}$ for definiteness) 
does not depend on $b_3$, but for another choice of $b_4, \dots , b_N$ the 
other vector $\ket{\beta}$ may be independent on $b_3$, while $\ket{\alpha}$ 
may depend on it. So, one cannot conclude that one of the vectors is 
\textit{generally} independent.

The problem is best illustrated with a counterexample. Consider the 
four-qubit
state
\be
\ket{\psi} = \frac{1}{2} ( \ket{0000} + \ket{0101} + \ket{0110} + \ket{1111}).
\ee
One can easily check that this is not a product state for any bipartition, 
so the state is genuine multiparticle entangled. Also, any projection into
the computational basis on the particles three and four leaves the first 
two particles in a product state. We have for the dependencies:
\be
\ket{\alpha(00)} = \ket{0}, \quad
\ket{\alpha(01)} = \ket{0}, \quad
\ket{\alpha(10)} = \ket{0}, \quad
\ket{\alpha(11)} = \ket{1},
\ee
and
\be
\ket{\beta(00)} = \ket{0}, \quad
\ket{\beta(01)} = \ket{1}, \quad
\ket{\beta(10)} = \ket{1}, \quad
\ket{\beta(11)} = \ket{1},
\ee
so neither of these vectors does depend on a single index only. 

Of course, if one chooses measurements in other directions on
the qubits three and four, that is, one measures vectors like
\be
\ket{c_3} = \cos(\gamma) \ket{0} + \sin(\gamma) \ket{1}
\mbox{ and }
\ket{c_4} = \cos(\delta) \ket{0} + \sin(\delta) \ket{1},
\ee
then the remaining state on the qubits one and two is entangled. So
the state $\ket{\psi}$ is not a counterexample to the main statement of
the Lemma, but it demonstrates that proof requires some extra work. 

Finally, if one accepts the step that each index [$b_i$] occurs only
in $\ket{\alpha}$ or $\ket{\beta}$, but not in both, one can conclude 
as demonstrated in Ref.~\cite{pr} that the original state has to 
factorize, so it is not entangled.

%%%%%%%%%%%%%%%%%%%%%%%%%%%%%%%%%%%%%%%%%%%%%%%
\section{Completing the argument}
%%%%%%%%%%%%%%%%%%%%%%%%%%%%%%%%%%%%%%%%%%%%%%%

The previous section demonstrated that the proof of the Lemma 
in Ref.~\cite{pr} is missing some discussions in order to be 
complete. In this section we  provide a way to add the missing 
part. We prove the following statement: 

{\it Let 
${\bf b'} =  (b_3, b_4,\dots, b_N)$ with $b_i\in\{0,1\}$  be the basis 
vectors which are used for the projection on the qubits $3, \dots, N$ 
and denote the remaining product state on the first two qubits by 
$\ket{\alpha(\bf{b'})}\ket{\beta(\bf{b'})}$. Then, $\ket{\alpha(\cdot)}$ 
depends only on some subset of the indices $\bf{b'}$, 
while $\ket{\beta(\cdot)}$ depends on the complement subset.} This statement
implies the correctness of the Lemma in Ref.~\cite{pr}.

The proof is done by assuming the opposite and reaching a contradiction.  
The opposite claim is that there exists an index $i$ (without the loss of 
generality, we can take $i=3$) and two sets of values for the remaining
indices  
\begin{equation}
\textbf{b}\; =b_4, b_5,\dots, b_N \quad \mbox{and} \quad  \textbf{B} \;=B_4, B_5,\dots, B_N,
\end{equation}
such that 
\begin{equation}
 \ket{\alpha(0,\textbf{b})}\neq\ket{\alpha(1,\textbf{b})}\quad 
 \mbox{\it and} 
 \quad\ket{\beta(0, \bf{B})}\neq\ket{\beta(1, \bf{B})},
\end{equation}
meaning that both depend on $b_3$.
Here, $\ket{\alpha(0,\textbf{b})}$ is a short-hand notation for 
$\ket{\alpha(b_3=0,\textbf{b})}.$ Also, the inequality symbol 
here and in the following indicates linear independence, i.e.,  
$\ket{\alpha(0,\textbf{b})}\neq \lambda\ket{ \alpha(1,\textbf{b})}$ 
for any $\lambda\neq 0.$  

The vectors $\textbf{b}$ and $\textbf{B}$ differ in some entries, 
but in some entries they match. Without  loss of generality, we 
can assume that they differ in the first $k$ entries while the 
others are the same and equal to zero. More specifically, they 
can be taken of the form:
\begin{align}
\mathbf{b} & = 0\ 0\ 0\ \dots\ 0\  \; 0\  0\ \dots\ 0,
\nonumber
\\
\mathbf{B}&=\underset{k}{\underbrace{1\ 1\ 1\ \dots\ 1}}\ \; \underset{N-k-3}{\underbrace{0\ 0\ \dots\ 0}}.
\end{align}
Then the  proof proceeds via induction on $k$. The precise statement we want to
prove for all $k$ is the following: Let the vectors $\mathbf{b}$ and 
$\mathbf{B}$ differ by at most at $k$ terms. Then, if $\ket{\alpha(0\mathbf{b})}\neq\ket{\alpha(1\mathbf{b})}$, the 
equality $\ket{\beta(0\mathbf{B})}=\ket{\beta(1\mathbf{B})}$ must 
hold. The crucial point here is that on each induction step we need 
to use the already derived linear dependencies and independencies 
from all the previous induction steps, i.e. for all $k' < k$.   
We give the first  $(k=0 \mapsto  k=1)$  and the second 
$(k=1\mapsto  k=2)$ step of the induction explicitly, as this 
is needed in order to get the idea for the general case. 
This general case is discussed afterwards. 

\begin{itemize}
%%%%%%%%%%%%%%%%%%%%%%
\item[(a)] The base case: If $k=0$, then $\textbf{b}=\textbf{B}$ and the 
proof for this particular case is included in the discussion in Section 2
and in Ref.~\cite{pr}.

%%%%%%%%%%%%%%%%%%%%%%
\item[(b)] $k=0 \mapsto k=1:$\\
As $\ket{\alpha(\cdot)}$ and $\ket{\beta(\cdot)}$  depend on $b_3$ and $b_4$ only, we 
can suppress the other indices and we write  $\ket{\alpha(00)}, \ket{\alpha(01)}$, etc.
Using this notation the problem boils down to showing that 
\begin{equation}
\label{step2}
\ket{\alpha(00)}\neq\ket{\alpha(10)} \quad \mbox{ and } \quad \ket{\beta(01)}\neq\ket{\beta(11)}
\end{equation}   
cannot happen simultaneously. We show that if this  would be true,  
it would contradict the assumption that the state 
after projection is a product state. 

{From} step (a) we already know that the statement is correct for $k=0$, which  
means that if only one value changes, then only one of $\ket{\alpha(\cdot)}$ 
and $\ket{\beta(\cdot)}$ can change.  
We can use this in the following way: For  $x=0$ or $1$,  
if $\ket{\alpha(0x)} \neq \ket{\alpha(1x)}$, it follows 
that $\ket{\beta(0x)}=\ket{\beta(1x)}$. Furthermore,
$\ket{\alpha(x0)} \neq \ket{\alpha(x1)} $ implies that 
$\ket{\beta(x0)}=\ket{\beta(x1)}$. 
And similarly, we can conclude equalities for the $\ket{\alpha(\cdot)}$ 
from inequalities of the $\ket{\beta(\cdot)}$.

Assuming that the statement in Eq.~(\ref{step2}) can be satisfied, we would like to 
reach a  contradiction. {From} the conditions in Eq.~(\ref{step2}) and our previous
argumentation it follows that 
\begin{equation}
\ket{\alpha(01)}=\ket{\alpha(11)} \quad \mbox{ and } \quad\ket{\beta(00)}=\ket{\beta(10)}.
\label{step2-a}
\end{equation} 
Now there are two cases to be considered and in both cases a contradiction is reached:

\begin{enumerate}
\item  The case $\ket{\alpha(00)}\neq\ket{\alpha(01)}$: \\
Then from the result for $k=0$ an equality  follows for the $\ket{\beta(\cdot)}$, 
namely $\ket{\beta(00)}=\ket{\beta(01)}.$
This implies with Eqs.~(\ref{step2-a}) and (\ref{step2}) that 
$\ket{\beta(10)} \neq \ket{\beta(11)}.$  
Consequently, we get that $\ket{\alpha(10)}=\ket{\alpha(11)}$
and from Eq.~(\ref{step2-a}) it follows that $\ket{\alpha(10)}=\ket{\alpha(01)}.$

To sum up all the relations for $\ket{\alpha(\cdot)}$ and $\ket{\beta(\cdot )}$, 
we can write:
\begin{align}
\ket{\tilde{\alpha}}\equiv\ket{\alpha(01)}=\ket{\alpha(10)}=\ket{\alpha(11)}\neq\ket{\alpha(00)},
\nonumber
\\
\ket{\tilde{\beta}}\equiv\ket{\beta(00)}=\ket{\beta(01)}=\ket{\beta(10)}\neq\ket{\beta(11)}.
\end{align}

Now we project the total state $\ket{\psi}$ onto 
$\ket{+}_3\ket{+}_4$  on the qubits 3 and 4 and on 
$\ket{0\dots0}_{5\dots N}$ on the further qubits.
Then, the remaining state  on the first two qubits
is:
\begin{equation}
\ket{\psi}_{12}
=
\sum_{i,j\in\{0,1\}}
\ket{\alpha(ij)}\ket{\beta(ij)}
=
\ket{\alpha(00)}\ket{\tilde{\beta}}+\ket{\tilde{\alpha}}\ket{\beta(11)}+2\ket{\tilde{\alpha}}\ket{\tilde{\beta}}.
\end{equation}
Noting that $\ket{\alpha(00)}\neq\ket{ \tilde{\alpha}}$, the state  $\ket{\psi}_{12}$ 
can only be a product state if 
\begin{equation}
\ket{\tilde{\beta}}=2\ket{\tilde{\beta}}+\ket{\beta(11)},
\end{equation}
which means that $\ket{\beta(11)}=\ket{\tilde{\beta}}$ up to a factor. 
This is a contradiction.

\item The complement case $\ket{\alpha(00)}=\ket{\alpha(01)}$:\\
Then $\ket{\alpha(11)}\neq\ket{\alpha(10)}$ and from 
Eqs.~(\ref{step2-a}) and (\ref{step2}) we have
\begin{equation}
\ket{{\alpha}} \equiv \ket{\alpha(00)}=\ket{\alpha(01)}=\ket{\alpha(11)} \neq 
\ket{\alpha(10)}.
\end{equation}
Furthermore, this implies that $\ket{\beta(10)}=\ket{\beta(11)}$. 
So,
\begin{equation}
\ket{{\beta}}=\ket{\beta(00)}=\ket{\beta(10)}=\ket{\beta(11)}\neq\ket{\beta(01)}.
\end{equation}
The proof then proceeds as in the case 1.
\end{enumerate}

%%%%%%%%%%%%%%%%%%%%%%%%%%%%

\item[(c)] $k=1 \mapsto k=2:$ As mentioned above, we also discuss this case in detail, as it
reveals the  main idea for the general case. 

This case also starts by assuming the opposite, similar to Eq.~(\ref{step2}):
\begin{equation}
\label{step3}
\ket{\alpha(000)}\neq\ket{\alpha(100)}
\quad \mbox{ and }\quad 
\ket{\beta(011)}\neq\ket{\beta(111)}.
\end{equation}
This means that if the $\bf{b}$ and ${\bf B}$ differ in two entries, we assume 
that it happens that $\ket{\alpha(\cdot)}$ and $\ket{\beta(\cdot)}$ both change
under a flip of the first index. 
%{From} the cases (a) and (b) we know 
%already  $\ket{\alpha(\cdot)}$ and $\ket{\beta(\cdot)}$ cannot both change, 
%if $\bf{b}$ and ${\bf B}$ differ in less than two entries. 
Eq.~(\ref{step3}) states that $\ket{\alpha(x00)}$ is not constant in $x$. Using 
the induction for $k=1$ it follows that $\ket{\beta(x00)}$, $\ket{\beta(x01)}$, 
and $\ket{\beta(x10)}$ are constant in $x$.
Consequently, from Eq.~(\ref{step3}) it follows that
\begin{align}
\label{alphaeq}
\ket{\alpha(001)}=\ket{\alpha(101)},\  \ket{\alpha(010)}=\ket{\alpha(110)},
\\
\label{betaeq} 
\ket{\beta(001)}=\ket{\beta(101)}, \ \ket{\beta(010)}=\ket{\beta(110)}.
\end{align}
Here we mention only  the relations that we use at this stage of the proof.

Throughout  the entire proof, to complete a particular induction step,  
it is crucial to deduce  many combinations of specific indices and  
flips for which  either  $\ket{\alpha(\cdot)}$ or $\ket{\beta(\cdot)}$ 
changes. These type of linear independencies are of a particular interest 
because from them it follows that for the same combination of indices and  
flips $\ket{\beta(\cdot)}$ or $\ket{\alpha(\cdot)}$ cannot change, 
respectively. Therefore,  for $\ket{\beta(\cdot)}$ here we consider the 
following single-index flips:
\begin{equation}
\ket{\beta(011)}\stackrel{F_3}{\longmapsto} \ket{\beta(010)} 
\stackrel{F_1}{\longmapsto} \ket{\beta(110)} \stackrel{F_3}{\longmapsto} \ket{\beta(111)},
\label{sequence1}
\end{equation}
where $F_i$ denotes a flip of the value of the $i$-th index.
Here the first and last term are linearly independent as assumed in Eq.~(\ref{step3}), but 
the second and third are equal according to Eq.~(\ref{betaeq}). So, since $\ket{\beta(\cdot)}$ 
changes from the first 
to the last term, it has to change either in the first step or the last one.\footnote{Also 
both can change, but there is no need to consider this separately as this is the simpler 
case: It includes the constraints from the both cases together. In fact, if both change, 
one even more directly see that this implies $\ket{\alpha(000)} = \ket{\alpha(100)}$
and a contradiction.} 
If $\ket{\beta(\cdot)}$ changes in the first step, $\ket{\beta(01x)}$ is not constant, 
implying that $\ket{\alpha(00x)}$, $\ket{\alpha(01x)}$, and $\ket{\alpha(11x)}$ are
constant in $x$.
In the other case, if $\ket{\beta(\cdot)}$ changes in the third step, $\ket{\beta(11x)}$ 
is not constant, implying that $\ket{\alpha(01x)}$, $\ket{\alpha(10x)}$, and 
$\ket{\alpha(11x)}$ are constant in $x$.

This implies that in any case, $\ket{\alpha(01x)}$ and $\ket{\alpha(11x)}$
have to be constant.
%$\ket{\alpha(\cdot)}$ cannot change under flips $F_3$ 
%of the third index for particular values of the other indices. Regardless
%whether $\ket{\beta(\cdot)}$ in Eq.~(\ref{sequence1}) changed in the first or 
%the last step, since $\ket{\beta(\cdot)}$ has to change in at least one 
%of the two flips it follows (with a brute force calculation) that  the 
%flips (F) of the last index while keeping the rest of the indices fixed, $(01F)$ 
%and  $(11F)$ cannot change $\ket{\alpha(\cdot)}$. To get these new equalities, 
%we used the results from the previous steps of induction as well and  
So, we have:
\begin{equation}
\ket{\alpha(011)} = \ket{\alpha(010)} 
\quad \mbox{ and } \quad  
\ket{\alpha(111)}=\ket{\alpha(110)}.
\end{equation}
For the future discussion  it is useful to summarize this finding by stating that 
a flip of the third index ($F_3$) cannot change $\ket{\alpha(\cdot)}$, unless the 
second index is zero.

The same argument can be applied to the different sequence:
\begin{equation}
\ket{\beta(011)}
\stackrel{F_2}{\longmapsto} \ket{\beta(001)} 
\stackrel{F_1}{\longmapsto} \ket{\beta(101)} 
\stackrel{F_2}{\longmapsto} \ket{\beta(111)}.
\end{equation}
Here again first and last term differ and the middle terms are equal.
Similar as above, this implies that the flips $(1F1)$ and $(0F1)$
cannot change $\ket{\alpha(\cdot)}$, i.e.
\begin{equation}
\ket{\alpha(101)}=\ket{\alpha(111)} \quad \mbox{ and } \quad   \ket{\alpha(001)}=\ket{\alpha(011)}.
\end{equation}
Again, we can state that $F_2$ cannot change $\ket{\alpha(\cdot)}$, unless the third 
index is zero.

{From} these observations it follows that:
\begin{equation}
\label{step3eq}
\ket{\alpha(001)}
\stackrel{F_2}{=}\ket{\alpha(011)}
\stackrel{F_3}{=}\ket{\alpha(010)}
\stackrel{\rm Eq. {\ref{alphaeq}}}{=}\ket{\alpha(110)}
\stackrel{F_3}{=}\ket{\alpha(111)}
\stackrel{F_2}{=}\ket{\alpha(101)}.
\end{equation}
This means that  $\ket{\alpha(\cdot)}$ can take only three values, 
two from the Eq.~(\ref{step3}) and one from the Eq.~(\ref{step3eq}).

Similarly, we can repeat the steps for the other side. Here, it turns our
that $\ket{\beta(\cdot)}$ cannot change under the flip $F_3$ (resp., $F_2$)
unless the second (resp., third) index is one. It follows that
$\ket{\beta(\cdot)}$ can take only three values as we have:
\begin{equation}
\ket{\beta(010)}\stackrel{F_2}{=}
\ket{\beta(000)}\stackrel{F_3}{=}
\ket{\beta(001)}\stackrel{\rm Eq. \ref{betaeq}}{=}
\ket{\beta(101)}\stackrel{F_3}{=}
\ket{\beta(100)}\stackrel{F_2}{=}\ket{\beta(110)}.
\end{equation}
There are again two cases to consider and each of them leads to a contradiction:

\begin{enumerate}

\item The case $\ket{\alpha(000)}\neq\ket{\alpha(001)}$:\\
Then from the induction requirement we have $\ket{\beta(011)}=\ket{\beta(010)}$, 
which implies that $\ket{\beta(\cdot)}$ can take only two values. Furthermore, from 
$\ket{\beta(111)} \neq \ket{\beta(110)}$ it follows that 
$\ket{\alpha(100)}=\ket{\alpha(101)}$, which itself means that $\ket{\alpha(\cdot)}$ 
can take only two values.\\
Then, we proceed as in the part (b) above. We project onto the product state  $\ket{+}_3\ket{+}_4\ket{+}_5\ket{0}^{\otimes N-5}$
and we cannot get a product state unless $\ket{\beta(\cdot)}$ is constant. This contradicts the initial 
assumption in Eq.~(\ref{step3}). 

\item The case $\ket{\alpha(000)}=\ket{\alpha(001)}$:\\
This means that $\ket{\alpha(\cdot)}$ can only take two values. Besides, $\ket{\alpha(101)}\neq\ket{\alpha(100)}$ 
implies that $\ket{\beta(111)}=\ket{\beta(110)}$. So,  $\ket{\beta(\cdot)}$ can take only two values. 
Similarly to the previous case, the projection onto the state
$\ket{+}_3\ket{+}_4\ket{+}_5\ket{0}^{\otimes N-5}$  leads to the contradiction.
\end{enumerate}

\item[(d)] The general case, $k-1\mapsto k:$

As before, we assume that the following inequalities happen simultaneously: 
\begin{equation}
\label{general}
\ket{ \alpha(0\ \underset{k}{\underbrace{00\dots0}})} \neq 
\ket{ \alpha(1\ \underset{k}{\underbrace{00\dots0}})}  
\mbox{ and } 
\ket{ \beta(0\ \underset{k}{\underbrace{11\dots1}})} \neq 
\ket{ \beta(1\ \underset{k}{\underbrace{11\dots1}})}. 
\end{equation}  
Similar to the previous cases, there are many equalities coming 
from the previous induction steps. Namely, we know already that a flip of
the value of one index cannot both change $\ket{\alpha(\cdot)}$
and $\ket{\beta(\cdot)}$, if the remaining indices differ at $k-1$
or less positions. This is the induction hypothesis that has to be used.

Now the following sequence of single index flips is considered:
\begin{equation}
\ket{\beta(01\dots11)}
\stackrel{F_{k+1}}{\longmapsto} \ket{\beta(01\dots10)} 
\stackrel{F_1}{\longmapsto} \ket{\beta(11\dots10)}
\stackrel{F_{k+1}}{\longmapsto} \ket{\beta(11\dots1)}.
\end{equation}
Here the first and the last terms  differ again according to 
the assumption, but the middle two entries are equal (this comes 
from the induction hypothesis mentioned above). So, $\ket{\beta(\cdot)}$ 
must change in the first or third step.

This implies that $\ket{\alpha(\cdot)}$ is not allowed to change under a flip 
$F_{k+1}$, unless the values of the indices $ 2, \dots, k$ are all zero. To see this, 
let us assume that they are not all zeros. This means that there is at least one 
digit in common with $1\dots1$ on the indices $2, \dots, k$.

On the one hand,  
then they differ only in $k-1$ or less digits from the string $01\dots 1$ on
the indices $1,\dots,k$. By the induction hypothesis we can conclude that 
$\ket{\alpha(\cdot)}$ is not allowed to change if $\ket{\beta(\cdot)}$ changes
in the first step of the sequence.

On the other hand, they also differ in less or equal $k-1$ digits from the string
$11 \dots $1 on the indices $1,\dots,k$, so $\ket{\alpha(\cdot)}$ is not allowed
to change if $\ket{\beta(\cdot)}$ changes in the third step of the sequence. Since 
$\ket{\beta(\cdot)}$ has to change in the first or third step, $\ket{\alpha(\cdot)}$ 
can not change under a flip $F_{k+1}$, unless the values of the indices $2,\dots,k$ 
are all zero.

By considering a similar sequence of flips, $F_{j}\circ F_1 \circ F_{j}$ with
$j \in \{2,\dots, k\}$, we can see that $\ket{\alpha(\cdot)}$ cannot change 
under a flip $F_j$, unless all of the indices $i \in \{2, \dots, k+1\}$ with 
$i \neq j$ are zero.

In addition, from the condition on $\ket{\beta(\cdot)}$ in Eq.~(\ref{general}) and the 
induction hypothesis it follows that 
\begin{equation}
\ket{\alpha(100\dots 01)}=\ket{\alpha(00\dots 01)}.
\label{alfafin}
\end{equation}
In combination with the invariance of $\ket{\alpha(\cdot)}$ under the 
flips $F_j$ mentioned above this implies that the terms in Eq.~(\ref{alfafin}) 
are also equal to all other $\ket{\alpha(\cdot)}$ except to the two
$\ket{\alpha(\cdot)}$ given in Eq.~(\ref{general}). Namely, any 
$\ket{\alpha(\cdot)}$ where not all indices $2,\dots,k+1$ are zero
can be brought to $\ket{\alpha(100\dots 01)}$ or $\ket{\alpha(00\dots 01)}$
by flips $F_j$ on the indices $j \in \{2,\dots, k+1\}$.

Therefore, $\ket{\alpha(\cdot)}$ can take only three values: two for the assumption 
in Eq.~(\ref{general}) and one for the rest of the ($2^{k+1}-2$ terms).  Exactly 
the same argument holds for $\ket{\beta(\cdot)}$. Consequently, we again consider 
two cases, reduce the three values to two for both $\ket{\alpha(\cdot)}$
and $\ket{\beta(\cdot)}$, and upon projecting on the product state, we reach 
the contradiction that $\ket{\beta(\cdot)}$ can only have one value. 
\end{itemize}
This ends the proof of the statement at the beginning.

In the following section, however, we need to consider 
the extension to the $d$-dimensional case. Besides, we 
need to be cautious that the projections in the computational
basis always give non-vanishing coefficients. Below we 
demonstrate that this condition can always be achieved 
and it does not impose any limitations on the Lemma.

\section{Additional Remarks}

\subsection{Generalization to the $d$-dimensional case} 

So far, we have discussed only multi-qubit systems and the question arises, whether
the results can be extended to higher dimensional systems. There are two ways to
deal with this issue. First, one can extend the discussion of Section 3 also to
the higher dimensions. The point is that in Section 3 the core conditions were always
stating that certain indices are equal or not equal. This, of course, can be formulated
also for non-binary indices. A second and more elegant way, however, makes use of the 
fact that any $N$-qudit entangled pure state can be projected by local means on an 
entangled $N$-qubit entangled state. This can be  achieved by the following procedure:

Let $\ket{\psi}$ be a $d \times d \times... \times d$ entangled state, where $d$ is the 
dimension of the Hilbert space associated with each system. We start by considering the 
first subsystem and write down the Schmidt decomposition with respect to the split 
$1|2,3,4,\dots,N$:
\begin{equation}
\ket{\psi} = \sum_{i=1}^d s_i \ket{i}_1 \ket{i}_{2,3,4,\dots,N}.   
\end{equation}
Then, we proceed as follows: 
\begin{enumerate}

\item If $\ket{\psi}$ is separable with respect to the $1|2,3,4,\dots,N$ 
partition, the sum consists only of one term and we do nothing. Note that 
effectively the whole state lives on a one-dimensional subspace on Alice's 
side, so one can view it as a $1 \times d  \times \dots \times d$ state.

\item If $\ket{\psi}$ is entangled with respect to the $1|2,3,4,\dots,N$  
partition, we project locally for Alice onto the two first Schmidt vectors, 
resulting in the truncated sum:
\begin{equation}
\ket{\psi'} = \sum_{i=1,2} s'_i \ket{i}_1 \ket{i}_{2,3,4,\dots,N}. 
\end{equation}
This state is still entangled with respect to the $1|2,3,4,\dots,N$ partition 
and it is a $2 \times d \times \dots \times d$ state.
\end{enumerate}

Now, we go on and do the same procedure iteratively for particle $2$, then particle $3$, 
etc., until we arrive at the last party $N$. At the end we have a state living in a 
$2 \times 2 \times \dots \times 2$-dimensional Hilbert space. If the original state was 
entangled, then the state $\ket{\psi'}$ is also entangled: when going through the parties, 
there will be some party, say $j$, where the last projection according to point (2) above 
is made. So the final state will be entangled with respect to the $j|{\rm \tt rest}$ partition 
(where ${\rm \tt rest}$ denotes all the remaining parties) and it is not a fully separable 
state. Note that the projection on $j$ may change the separability properties of the $1|{\rm \tt
rest}$ partition, and this partition 
may become separable. However, at least one entangled partition remains, and as 
discussed above, this is sufficient to show that the state is not a fully 
separable state. 

So any entangled $d \times d \times \dots \times d$ pure state can be projected 
locally onto a pure entangled $N$-qubit state and for qubit states we already 
repaired the proof.

\subsection{The question of non-vanishing coefficients}

A further point worth to discuss is the question whether the state $\ket{\psi}$
considered in the proof has maybe vanishing coefficients in the basis where the 
projections are made. In fact, the careful reader may have noticed that for the 
theorem and  the proof above to work, it is  required all possible projections 
in the computational basis, especially the states $\ket{\alpha(\cdot)}$ and 
$\ket{\beta(\cdot)}$ are non-zero. This might not be fulfilled for a given 
basis. 

Of course, for a random choice of the product basis this will be in general 
fulfilled. More constructively, one can ask whether there is a set of local 
unitaries that, when applied to any initial state $\ket{\psi}$ in the 
computational basis, give {\it with certainty} some states where all the 
coefficients are non-vanishing. Interestingly, this question was brought 
up as one of the ``ten most annoying questions in quantum computing'' 
\cite{Scott} with the local unitaries being the Hadamard gates and the 
solution was given in Ref.~\cite{Montamaro}. We now recall this result 
in the following lemma with the notation $\sigma_x$ and $\sigma_{z}$ 
being the Pauli-$X$ and Pauli-$Z$ matrices, respectively.

{\bf Lemma \cite{Montamaro}.} 
{\it Given an $N$-qubit pure state, 
there is always a way to apply Hadamard gates to some subset 
of the qubits to make all $2^N$ computational basis components 
having non-zero amplitudes. In other words, if one considers the 
$2^N$ product bases defined by the eigenstates of the observables 
$\sigma_{k_1}^{(1)} \otimes \dots \otimes \sigma_{k_N}^{(N)}$
with $\sigma_k^{(j)} \in \{\sigma_x, \sigma_z\}$, then any state 
$\ket{\psi}$ has non-vanishing coefficients in at least one of 
these bases.}

This Lemma guarantees that a suitable basis can be found in a constructive manner.
\qed

\subsection*{Acknowledgments}
We thank Carmen Constatin and Matty Hoban for discussions. We especially 
thank Sandu Popescu, Daniel Rohrlich, and Paul Skrzypczyk for comments on 
the manuscript, they agree on the points raised here. This work was 
supported by the DFG, the ERC (Consolidator Grant 
683107/TempoQ) and the FQXi Fund (Silicon 
Valley Community Foundation).


\begin{thebibliography}{15}
\bibitem{pr}
S. Popescu and D. Rohrlich, Phys. Lett. A {\bf 166}, 293 (1992).

\bibitem{gisin} 
N. Gisin and A. Peres, Phys. Lett. A {\bf 162}, 15 (1992).


\bibitem{Scott} S. Aaronson, {\it The ten most annoying questions in quantum computing} 
(2006), available at {\tt www.scottaaronson.com/blog/?p=112}.

\bibitem{Montamaro} 
A. Montanaro and D. J. Shepherd, 
{\it Hadamard gates and amplitudes of computational basis states} (2006),
available at {\tt www.scottaaronson.com/hadamard.pdf}.


\end{thebibliography}
\end{document}